\documentclass[useAMS,usenatbib,letter]{mn2e}

\usepackage{graphicx,url,color,rotating,amssymb,amsmath,hyperref}
\everymath{\displaystyle}
\usepackage{dcolumn}
\usepackage{bm}
\newcommand{\beq}{\begin{equation}}
\newcommand{\eeq}{\end{equation}}
\newcommand{\bal}{\begin{aligned}}
\newcommand{\eal}{\end{aligned}}
\newcommand{\beqa}{\begin{eqnarray}}
\newcommand{\eeqa}{\end{eqnarray}}

\newcommand{\Msun}{\rm M_\odot}
\newcommand{\hiMsun}{h^{-1} \rm M_\odot}
\newcommand{\hiGpc}{h^{-1} \rm Gpc}
\newcommand{\hiMpc}{h^{-1} \rm Mpc}
\newcommand{\hikpc}{h^{-1} \rm kpc}

\newcommand{\kms}{{\rm km\, s^{-1}}}

\newcommand{\Mvir}{M_{\rm vir}}
\newcommand{\Rvir}{R_{\rm vir}}
\newcommand{\sigmavir}{\sigma_{\rm vir}}
\newcommand{\sigmaDM}{\sigma_{\rm DM}}
\newcommand{\sigmagal}{\sigma_{\rm gal}}

\newcommand{\vmax}{v_{\rm max}}
\newcommand{\vpk}{v_{\rm pk}}

\newcommand{\vtoday}{v_{\rm 0}}
\newcommand{\Mstar}{M_{\rm star}}
\newcommand{\bv}{b_{\rm v}}

\newcommand{\DMsub}{N\mbox{-}{\rm body}\mbox{-}v_0}
\newcommand{\DMgal}{N\mbox{-}{\rm body}\mbox{-}v_{\rm pk}}
\newcommand{\HDsub}{{\rm Hydro}\mbox{-}v_0}
\newcommand{\HDgal}{{\rm Hydro}\mbox{-}M_{\rm star}}

\newcommand{\comment}[1]{}

\begin{document}
\title[Velocity bias]{Virial scaling of galaxies in clusters: bright to faint is cool to hot}
\author[H.-Y.~Wu et al.]{
Hao-Yi Wu,$^{1}$\thanks{E-mail: hywu@umich.edu}
Oliver Hahn,$^{2}$ 
August E. Evrard,$^{1}$
Risa H. Wechsler,$^{3}$ 
Klaus Dolag$^{4,5}$\\
$^{1}$ Department of Physics, University of Michigan, Ann Arbor, MI 48109, USA\\  
$^{2}$ Department of Physics, ETH Zurich, CH-8093 Z\"urich, Switzerland \\
$^{3}$ KIPAC, Stanford University, SLAC National Accelerator Laboratory, Menlo Park, CA 94025, USA\\ 
$^{4}$ Universit\"atssterenwarte, D-81679 M\"unchen, Germany\\  
$^{5}$ Max-Planck-Institut f\"ur Astrophysik, D-85748 Garching, Germany\\
}

\date{\today}

\maketitle

\begin{abstract}
By combining galaxy tracers from high-resolution $N$-body and
hydrodynamical simulations, we present a consistent picture of the
behaviour of galaxy velocities in massive clusters.  In haloes above
$\sim \!  10^{14} \Msun$, the brightest satellite galaxies are
slightly cooler compared to the dark matter, while fainter satellites
are hotter.  Within the virial radius of a cluster, the mean velocity
dispersion based on the 100 brightest galaxies is a factor of $1.065
\pm 0.005 \, \rm{(stat)} \, \pm 0.027 \ \rm{(sys)}$ higher than that
of the dark matter (corresponding to a $\sim \! 10-15$ per cent bias in
the dynamical mass estimate) while that based on only the five
brightest galaxies is $0.868 \pm 0.039 \, \rm{(stat)} \, \pm 0.035\,
\rm{(sys)}$.  These trends are approximately independent of redshift.
The velocity structure is sensitive to the modelling of galaxies in
clusters, indicative of the complex interplay of tidal stripping,
dynamical friction, and merging.  Velocity dispersions derived from
instantaneous subhalo properties are larger than those employing
either peak subhalo properties or hydrodynamical galaxy tracers.  The
latter two methods are consistent, implying that stacked spectroscopic
analysis of cluster samples should, after correction for projection,
show a trend towards slightly higher velocities when fainter galaxies
are included, with an unbiased measure of dark matter velocity
dispersion coming from approximately 30 galaxies per cluster.  We show
evidence that the velocity distribution function of bright galaxies
near the cluster centre has a low-velocity tail due to strong
dynamical friction.
\end{abstract}


\section{Introduction}

Dynamical measurements of galaxy cluster masses --- based on the
line-of-sight galaxy velocities from spectroscopic observations ---
have a long history, using a broad range of methods, including the
direct application of the virial theorem
\citep[e.g.,][]{YahilVidal77,Danese80,Heisler85,Girardi93,Lubin93,Carlberg97},
the caustic method measuring the escape velocity of galaxies
\citep[e.g.,][]{DiaferioGeller97,Diaferio99}, and Jeans analysis
\citep[e.g,][]{LokasMamon03,Mamon13}.  Recent cluster surveys have
enabled new studies of the relation between velocity dispersion and
other cluster mass proxies, including optical richness (e.g.,
\citealt{Becker07}, and \citealt{AndreonHurn10} based on SDSS),
Sunyaev-Zel'dovich effect (e.g., \citealt{Rines10} based on HeCS, and
\citealt{Sifon12} based on ACT), and X-ray luminosity or temperature
(e.g., \citealt{Zhang11} based on HIFLUGCS, and \citealt{Rines13}
based on HeCS).  With the current observational resources, most of
these studies focus on a few tens of clusters with $\sim$100 member
galaxies for each cluster, i.e., $\sim$10,000 spectra in total.  These
samples are often limited by statistics; upcoming deep, wide
spectroscopic surveys such as {\em Euclid} \citep{Laureijs11}, {\em
Subaru} Prime Focus Spectrograph \citep{Ellis12}, and
BigBOSS\footnote{\tt http://bigboss.lbl.gov} will allow more detailed
study of how dynamical masses relate to other mass proxies.

On the theoretical side, the virial scaling relation between halo mass
and the velocity dispersion of {\em dark matter} (DM) has been
established at high precision using simulation ensembles
\citep{Evrard08}.  The dynamics of galaxies, however, may be different
from that of the DM due to a number of physical processes, such as
dynamical friction, tidal stripping and disruption, mergers, and
hydrodynamic drag by the gas in the cluster.  This ``velocity bias''
of galaxies with respect to DM has been explored with ever improving
simulations and cluster galaxy models
\cite[e.g.,][]{Carlberg90,Carlberg94,Evrard94,Summers95,Frenk96,Ghigna00,Colin00,Gill04b,Gao04,Diemand04,Faltenbacher06,Lau10,Munari13}.
However, discrepancies in the measured velocity bias exist among
simulations with different numerical treatments and different galaxy
tracers (see Section~\ref{sec:rev} for a review).  A principal goal of
this paper is to demonstrate the origin of these differences and to
show that a consistent picture appears to be emerging when
state-of-the-art galaxy tracers are used.

In this work, we focus on the velocity dispersion of galaxies in
massive clusters, using the $N$-body simulation suite {\sc rhapsody}
and the hydrodynamical simulation {\sc magneticum pathfinder}.  We
study the velocity bias using various galaxy tracers in both
simulations and compare with results in the literature.  Using the
rich galaxy statistics from {\sc rhapsody}, we quantify statistically
the virial scaling relation obtained from different galaxy
populations.  We then compare the dynamical structure of galaxies in
both simulations.

We note that the 3D effects presented in this work will be complicated
by the projection effect in observations, which both increases scatter
and introduces potential biases in cluster mass estimates
\citep[e.g.,][]{Biviano06,Cohn07,Wojtak07,Mamon10,WhiteM10,Cohn12,Saro12,Gifford13,Gifford13b}.
Here, we focus on quantifying the theoretical uncertainty in the {\em
intrinsic} velocity structure of clusters in order to pave the way for
future studies that incorporate observational complications.

This paper is organized as follows.  Section~\ref{sec:sims} introduces
the simulation data and the galaxy tracers used in this work.
Section~\ref{sec:rev} compares the velocity bias results in this work
and in the literature and presents a converging picture.
Section~\ref{sec:rank} discusses the virial scaling calibrated from
galaxies.  In Section~\ref{sec:bv_dm}, we explore in detail the
velocity structure of cluster galaxies to better understand the
physical processes causing velocity bias.  In
Section~\ref{sec:bv_hydro}, we compare the velocity structures in
$N$-body and hydrodynamical simulations.  We conclude in
Section~\ref{sec:summary}.

\section{Simulations}\label{sec:sims}
\begin{table*}
\centering
\begin{tabular}{
>{\centering}p{1.2cm}<{\centering}  
>{\centering}p{2cm}<{\centering} 
>{\centering}p{2cm}<{\centering} 
>{\centering}p{2cm}<{\centering} 
>{\centering}p{2.2cm}<{\centering} 
>{\centering}p{2cm}<{\centering} 
c
}
\hline
{Type} &
{Name} &
{Mass res. ($10^{8}\hiMsun$)} &
{Force res. ($\hikpc$)} & 
$\rm N_{halo}$&
$\rm M_{halo}$ ($10^{14}\hiMsun$) & \\
\hline
$N$-body& 
{\sc rhapsody} &
1.3 & 
3.3 &  
96 ($z$ = 0)\\ 
8142 ($0\leq z\leq2$)&
$6.3\pm0.48$ \\  
$2.85\pm1.99$ &
\\
\hline
Hydro& 
{\sc magneticum pathfinder}&
6.9 (DM)\\ 1.4 (gas)& 
5&  
46 ($z$ = 0)&
$>1$
&\\
\hline
\end{tabular}
\caption{Simulations used in this work.}
\label{tab:sims}
\end{table*}

In this work, we use a suite of zoom-in DM simulations of galaxy
clusters called {\sc rhapsody} and a cosmological box from the
hydrodynamical simulation {\sc magneticum pathfinder}.  The simulation
parameters are summarized in Table~\ref{tab:sims}.  Although these two
sets of simulations are based on slightly different cosmological
parameters (see below), we expect that the impact on our results is
negligible because \cite{Evrard08} have demonstrated that the virial
scaling is insensitive to cosmology.

\subsection{Pure dark matter $N$-body simulation: Rhapsody}

The {\sc rhapsody} simulation (\citeauthor{Wu12} 2013a) is a
statistical sample of high-resolution $N$-body re-simulations of
galaxy clusters, designed for studying the scatter of the
observable--mass relation, a property essential for precision
cosmology.  The main sample includes 96 haloes at $z=0$ with virial
masses lying in a narrow range, $\rm log_{10}\Mvir = 14.8\pm 0.05$,
re-simulated from a $1 \,\hiGpc$ volume.  The mass resolution is
$1.3\times10^8 \,\hiMsun$ (equivalent to $8192^3$ particles in this
volume), and the gravitational softening length is $3.3\,\hikpc$.  In
addition to haloes at $z=0$, we include 8142 most-massive progenitors
of them at 85 discrete redshifts up to $z = 2$ to calibrate the viral
scaling relation.

The {\sc rhapsody} simulation is based on the {\sc music} code
\citep{HahnAbel11} to generate the multi-scale initial condition, the
public version of {\sc gadget-2} \citep{Springel05} to perform the
gravitational evolution, the {\sc rockstar} (\citeauthor{Behroozi11rs}
2013a) phase-space halo finder to identify haloes and subhaloes, and
the gravitationally consistent merger tree code
(\citeauthor{Behroozi11tree} 2013b) to track the evolution history for
haloes and subhaloes.  The key properties of the main haloes and
subhaloes in this sample have been presented in \citeauthor{Wu12}
(2013a) and \citeauthor{Wu12b} (2013b), respectively.

The cosmological parameters used in {\sc rhapsody} are based on a flat
$\Lambda$ cold dark matter cosmology: 
$\Omega_{\rm M}$ = 0.25, 
$\Omega_{\rm b}$ = 0.04, 
$h$ = 0.7, 
$\sigma_8$ = 0.8, 
and $n_s$ = 1.

\subsection{Hydrodynamical simulation: Magneticum Pathfinder}

The {\sc magneticum pathfinder}\footnote{\tt
http://www.mpa-garching.mpg.de/$\sim$kdolag/Simulations/} is a new
suite of hydrodynamical simulations that aim for multi-wavelength
studies of galaxy clusters (Dolag et al.~in preparation).  This work
is based on Box 3 of that suite, a cosmological volume with a side
length $128\,\hiMpc$ and $2\times 576^3$ particles.  The mass of DM
particles is $6.9\times10^8 \,\hiMsun$, and the mass of SPH gas
particles is $1.4\times10^8 \,\hiMsun$.  The gravitational softening
length is $5\,\hikpc$.  This work uses the 46 haloes with $\Mvir >
10^{14}\,\hiMsun$ in this volume.

{\sc magneticum pathfinder} uses the smoothed particle
magnetohydrodynamics code {\sc p-gadget3 (xxl)}, which includes star
formation and chemical enrichment, active galactic nuclei (AGN)
feedback, thermal conduction, passive magnetic fields, and magnetic
dissipation.  The subhaloes and galaxies in this simulation are
identified with {\sc l-subfind}, presented in \cite{Dolag09}.  In
brief, {\sc l-subfind} improves upon the original {\sc subfind} code
\citep{Springel01} to apply to DM, stellar, and gas particles.  It
identifies friends-of-friends structures based on DM particles,
associates star and gas particles to the nearest DM particles, and
calculates the SPH kernel density for each species separately.  The
unbinding procedure takes into account the internal thermal energy of
gas particles.  The centres of subhaloes are chosen to be at the local
minimum of the gravitational potential.

The cosmological parameters used in {\sc magneticum pathfinder} are
based on {\em Wilkinson Microwave Anisotropy Probe} 7
\citep{Komatsu11}:
$\Omega_{\rm M}$ = 0.272, 
$\Omega_{\rm b}$ = 0.0456,
$\Omega_\Lambda$ = 0.728,
$h$ = 0.704,
$\sigma_8$ = 0.809, and 
$n_s$ = 0.963.

\subsection{Galaxy tracers in simulations}

We consider a rank-ordered list of subhaloes in a halo as a proxy for
a rank-ordered list of cluster galaxies.  For the {\sc rhapsody}
simulation, subhaloes are ranked by their maximum circular velocity of
DM, $\vmax = {\rm max}[\sqrt{{G M(<r)}/{r}}]$, at two different
epochs: an instantaneous value, $\vtoday$, which denotes the value of
$\vmax$ evaluated at $z=0$; $\vpk$, the peak value of $\vmax$ over its
prior history.  Ranking subhaloes by $\vpk$ is known to be superior to
$\vtoday$ ranking when comparing to mass- or luminosity- selected
samples of galaxies; for example, recent analysis by \cite{Reddick13}
has shown that the $\vpk$ prescription reproduces observed galaxy
clustering in the local Universe with remarkable fidelity \citep[also
see, e.g.,][]{NagaiKravtsov05,Conroy06, Wetzel12}.

For the {\sc magneticum pathfinder} simulation, we employ both
$\vtoday$ (derived from all mass components) and the stellar mass in
subhaloes at $z = 0$ for ranking galaxies.  Because cooling and star
formation creates a more core-dominated subhalo structure in
hydrodynamical simulations, we will see that using $\vtoday$ in {\sc
magneticum pathfinder} produces results consistent with those using
the stellar mass, and also consistent with those using $\vpk$ in {\sc
rhapsody}.

Our notation and a summary of these galaxy tracers are given in Table~\ref{tab:bvDefn}.  

\begin{table}
\centering
\setlength{\tabcolsep}{0.5em}
\begin{tabular}{
>{\centering}p{2cm}<{\centering}  
>{\centering}p{4.5cm}<{\centering}  
c
}
\hline
\rule[-2mm]{0mm}{6mm} Label &  Proxy of  & \\ \hline
\rule[-2mm]{0mm}{6mm} $\DMsub$      &  Subhalo mass in $N$-body sims        &   \\ \hline
\rule[-2mm]{0mm}{6mm} $\DMgal$       &  Galaxy luminosity in $N$-body sims    & \\ \hline
\rule[-2mm]{0mm}{6mm} $\HDsub$      &   Subhalo mass in hydro sims       &  \\ \hline
\rule[-2mm]{0mm}{6mm} $\HDgal$       &  Galaxy stellar mass in hydro sims  &  \\ \hline
\end{tabular}
\caption{Summary of galaxy tracer populations.
}
\label{tab:bvDefn}
\end{table}

\section{Velocity bias: a converging picture}\label{sec:rev}
\begin{figure*}
\includegraphics[width=1.5\columnwidth]{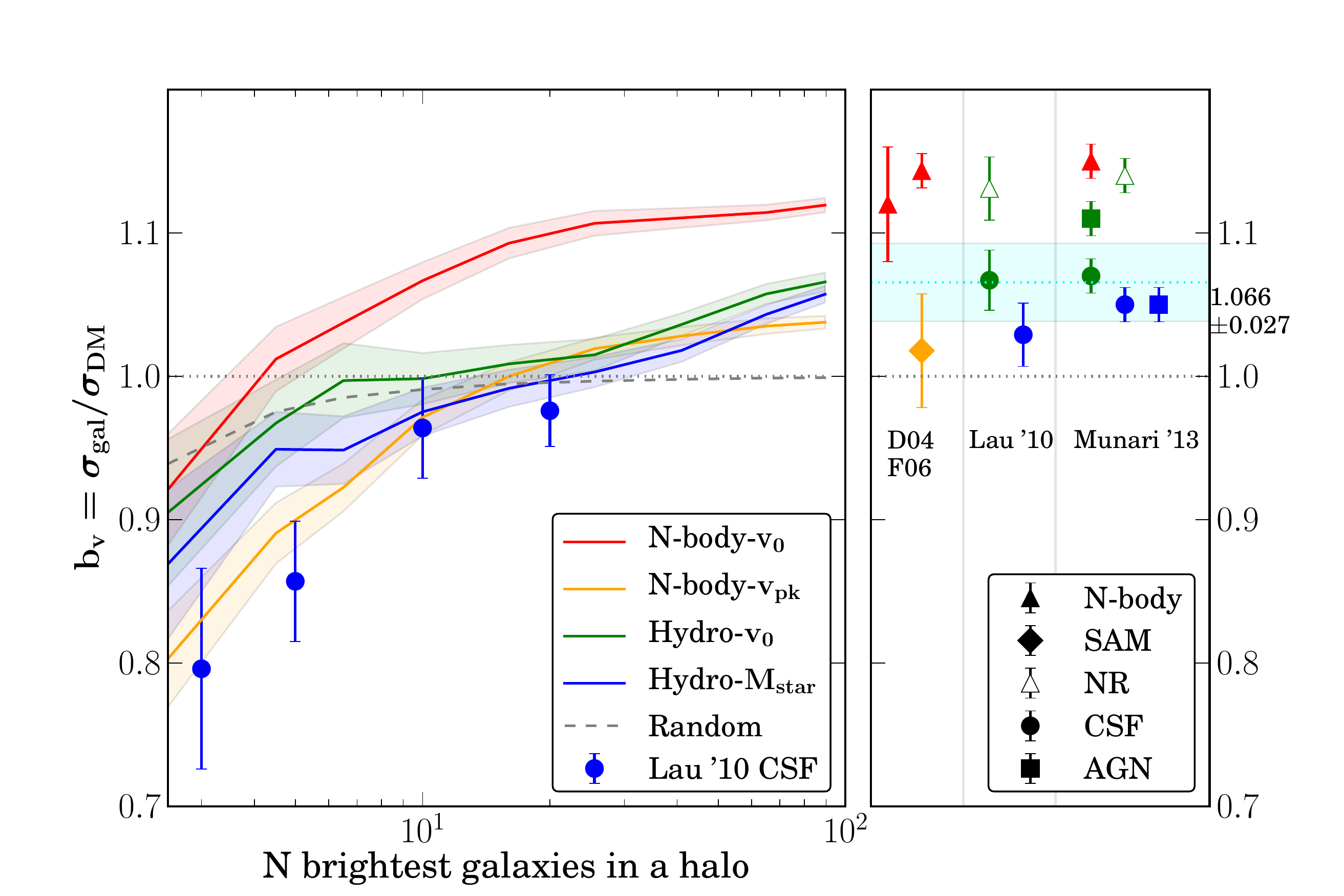}
\caption{ Velocity bias based on various simulations and galaxy
tracers.  The curves are the results of this work, showing how
velocity bias depends on the number of brightest galaxies used to
estimate the velocity dispersion.  Subhaloes in $N$-body simulations
(red) show substantial velocity bias, while galaxies modelled from
$N$-body and hydro simulations (orange and blue) have a smaller
velocity bias.  Using $\vpk$ as a luminosity proxy in the $N$-body
simulation gives similar results as using $\vtoday$ and $\Mstar$ in
the hydro simulation.  }
\label{fig:rank}
\end{figure*}
\begin{table}
\centering
\setlength{\tabcolsep}{0.5em}
\begin{tabular}{
>{\centering}p{2cm}<{\centering}  
>{\centering}p{2cm}<{\centering}  
>{\centering}p{2cm}<{\centering} 
c
}
\hline
\rule[-2mm]{0mm}{6mm} Tracer &   $\bv$ for $N=5$ & $\bv$ for $N=100$  & \\ \hline
\rule[-2mm]{0mm}{6mm} $\DMsub$ & 1.018$\pm$0.013 & 1.119$\pm$0.005 & \\ \hline
\rule[-2mm]{0mm}{6mm} $\DMgal$ & 0.899$\pm$0.012 & 1.038$\pm$0.004 & \\ \hline
\rule[-2mm]{0mm}{6mm} $\HDsub$ & 0.975$\pm$0.018 & 1.066$\pm$0.007 & \\ \hline
\rule[-2mm]{0mm}{6mm} $\HDgal$ & 0.949$\pm$0.017 & 1.057$\pm$0.006 & \\ \hline
\rule[-2mm]{0mm}{6mm} Random & 0.978 & 0.999 & \\ \hline
\end{tabular}
\caption{Mean velocity bias of different galaxy tracer populations. 
}
\label{tab:bv}
\end{table}

The 3D velocity dispersion of a group of $N$ objects (DM particles or galaxies)
in a system can be defined as
\beq
\sigma^2 = \frac{1}{N-1}\sum_{i=1}^{N} || {\bm v}_i - \bar{\bm v} ||^2 \ ,
\label{eq:sigma}
\eeq
where ${\bm v}_i$ denotes the 3D velocity of the $i$th object, and
$\bar{\bm v}$ denotes the mean velocity of these $N$ objects.

This difference between the velocity dispersion of galaxies
($\sigmagal$) and that of DM ($\sigmaDM$) can be quantified by {\it
velocity bias}, defined as:
\beq
\bv = \frac{\sigmagal}{\sigmaDM} \ .
\eeq
When we calculate $\sigmaDM$, we use all particles within $\Rvir$
around the central halo, which also corresponds to the brightest
cluster galaxy (BCG) in our modelling.  When we calculate $\sigmagal$,
we use all galaxies {\em except for the BCG} within $\Rvir$.  Here
$\Rvir$ is the radius encompassing the overdensity $\Delta_{\rm vir}$
calculated from a spherical collapse model \citep{GunnGott72}.  We
employ the fitting function of \cite{BryanNorman98}.

In each calculation, $\bar{\bm v}$ corresponds to the mean velocity of
the particular population in question rather than the bulk velocity of
the halo or the velocity of the BCG.  This choice gives the minimal
velocity dispersion and is presumably the closest to reality.

\subsection{Velocity bias of different galaxy tracers}

Fig.~\ref{fig:rank} compares our results of velocity bias (left-hand
panel) with results from the literature (right-hand panel).  Let us
first focus on the left-hand panel, which presents $\bv$ as a function
of $N$, the number of brightest galaxies or most massive subhaloes
used to measure $\sigmagal$.  The red/orange curve corresponds to
$\DMsub$/$\DMgal$, while the green/blue curve corresponds to
$\HDsub$/$\HDgal$.  We present the mean $\bv$ of the $z=0$ sample and
the error on the mean.  The grey dashed curve corresponds to selecting
$N$ random DM particles for measuring the velocity dispersion,
reflecting the bias inherent to measuring $\sigma$ with finite $N$.

When the number of galaxies is low, the velocity bias is less than 1
and slightly below the value expected from randomly sampled particles
(although $\DMsub$ and $\HDsub$ are consistent with the random
case). This indicates that the brightest galaxies tend to be slower
than DM particles and suggests a strong influence of dynamical
friction (see e.g.\ \citealt{Goto05} for observational evidence).  On
the other hand, when the number of galaxies is high, the velocity bias
becomes greater than 1. This can be explained by the inclusion of
fainter galaxies/low-mass subhaloes into the sample.  The slow faint
galaxies tend to be tidally disrupted or merge with the BCG, so that
the surviving galaxies tend to be a fast population, thus biasing the
velocity dispersion.  We will further investigate how velocity bias
depends on galaxy luminosity in Section~\ref{sec:bv_dm}.

We note that using $\DMsub$ tends to overestimate $\bv$, indicating
that by using $\vtoday$ the sample of kinematic tracers becomes biased
since strongly stripped, early accreted subhaloes drop out of the
sample.  These drop-outs tend to be slow, thus the velocity dispersion
is increased.  Comparing $\vtoday$ in both cases, we note that
$\HDsub$ has lower velocity bias, indicating that subhaloes are less
prone to stripping and disruption in the hydro simulation than in the
DM simulation.  In fact, $\HDsub$ behaves more similar to $\DMgal$ at
high $N$.  Although the clusters in $N$-body and hydro simulations
have different masses (as indicated in Table~\ref{tab:sims}), we note
that the results here are roughly independent of halo mass and
redshift, which will be shown in Fig.~\ref{fig:redshift}.  In
Table~\ref{tab:bv}, we compare $\bv$ from these four galaxy/subhalo
populations.

\subsection{Comparison with the results in the literature}

The right-hand panel of Fig.~\ref{fig:rank} shows the comparison
between our results and the following results from the literature: (1)
\citet[][D04]{Diemand04} and \citet[][F06]{Faltenbacher06}; (2)
\cite{Lau10}; and (3) \cite{Munari13}.  These results are
representative of the diversity of the predictions from different
simulations, which can usually be put into one of these four
categories.
\begin{enumerate}

\item {\em Subhaloes from dark matter $N$-body simulations} (marked as
red).  The velocities of ``resolved'' subhaloes are used to compute
the velocity dispersion, and subhaloes are selected based on their
current mass or $\vtoday$ (like our $\DMsub$).  These samples show
$\bv > 1$, but the value of $\bv$ depends on the resolution of the
simulation and the criterion of subhalo selection
\citep[e.g.,][]{Colin00,Diemand04,Faltenbacher06}.  Increasing the
resolution tends to reduce the velocity bias, indicating that the
positive velocity bias can be a result of over-merging, which tends to
remove slow and less massive subhaloes.

\item {\em Galaxies inferred from dark matter $N$-body simulations}
(marked as orange).  There are two common methods to predict the
galaxy properties from DM $N$-body simulations.  The first method is
entirely based on resolved subhaloes and assigning galaxy luminosity
based on the subhalo mass before accretion (like our $\DMgal$).  We
find that $\DMgal$ leads to smaller $\bv$ than $\DMsub$, indicating
that the most massive subhaloes and the brightest galaxies in a
cluster are two populations with different dynamical properties.  The
former population tends to exclude the highly stripped subhaloes,
which have been accreted early and have slow velocities, and include
the recently accreted subhaloes, which still have high $\vtoday$ and
high orbital velocities \citep[e.g.,][]{Gao04}.  The second method is
to apply semi-analytic models (SAMs) for galaxy formation and to track
galaxies using their ``most-bound particles'' after their subhaloes
are disrupted.  In this case, $\bv$ is greatly reduced because a
fraction of the galaxies, by construction, have the same velocities as
DM particles.  For example, \cite{Faltenbacher06} have found such a
trend between their two samples based on subhaloes and most-bound
particles (red and orange triangles; also see, e.g.,
\citealt{Gao04,Faltenbacher05,Sales07a}).  While using this method,
one should caution that SAMs do not always reproduce the observed
galaxy number density profile at small radii
\citep[e.g.,][]{Budzynski12}.

\item {\em Subhaloes in hydrodynamical simulations} (marked as green).
This is mainly used to compare with category (i) to illustrate the
impact of baryonic physics.  We use resolved subhaloes in the hydro
simulation based on the mass proxy $\vtoday$ ($\HDsub$), and its $\bv$
is lower than $\DMsub$ but similar to $\DMgal$.  This indicates that
subhaloes in the hydro simulation are more resistant to stripping than
their counterparts in $N$-body simulations, because of the star
particles condensed in the centre of subhaloes.  This result also
implies that $\vtoday$ in the hydro simulation behaves similar to
$\vpk$ in the DM simulation.  However, simulations with different
baryonic physics have reported various values.  In the right-hand
panel, we show various results from \cite{Lau10} and \cite{Munari13},
which includes simulations with ``NR'' (non-radiative, open
triangles), ``CSF'' (cooling and star formation, circles), and ``AGN''
(feedback from active galactic nuclei, squares).  In general, ``NR''
processes tend to give results very similar to pure DM simulations
because these simulations tend to produce puffy subhaloes which are
prone to stripping; ``CSF'' can reduce the velocity bias because it
can produce stellar cores; ``AGN'' again brings up velocity bias
because the core density is reduced.  The results could still vary
based on the implementation of the feedback.  We note that these
different feedback processes can also alter the density profile of the
main halo and thus $\sigmaDM$; for example, CSF tends to generate a
high-density stellar core, while AGN tends to reduce the core density
\citep[e.g.,][]{Dubois11,Martizzi12}.

\item {\em Galaxies in hydrodynamical simulations} (marked as blue).
Hydrodynamical simulations with star formation prescriptions can
directly predict the properties of galaxies.  These galaxies have been
consistently shown to have very small velocity bias.  We note that
$\bv$ from $\HDgal$ is slightly smaller than that of $\HDsub$.  We
also show results from \citet[][CSF]{Lau10} and \citet[][CSF and
AGN]{Munari13} as blue points.  However, it is still computationally
prohibitive to generate a statistical sample of high-resolution
cluster galaxies with an adequate span of the range of plausible
recipes for baryonic physics to minimize both the statistical and
systematic errors.

\end{enumerate}

Fig.~\ref{fig:rank} shows a converging picture for the velocity bias
obtained from simulations based on different techniques.  Combing our
measurements from $\DMgal$, $\HDsub$, and $\HDgal$, as well as the
$\bv$ values from SAM, CSF, and AGN in the literature [i.e. the
orange, blue, and green points (excluding NR) and curves, from which
we take the value at high $N$] leads to $1.065 \pm 0.005 \,
\rm{(stat)} \, \pm 0.027 \ \rm{(sys)}$, where the statistical error is
the error on the weighted mean, while the systematic error is the
weighted sample variance among different measurements.  The same
calculation for $N=5$ gives $0.868 \pm 0.039 \, \rm{(stat)} \, \pm
0.035\, \rm{(sys)}$.

While we were finalizing this manuscript we learned of the related
work by \citet{Old13}.  They use a semi-analytic treatment to trace
galaxies and find values of $\bv$ below unity at the bright end that
are consistent with our findings.  They also see $\bv$ increasing for
faint galaxies, but do not find values $\bv$ above unity.  There are
subtle differences between their analysis methods and ours which may
be driving the faint-end differences.  We note that: (1) the
hydrodynamical simulations used here, which one can argue are the most
realistic of the methods under discussion, show $\bv = 1.06$ at
$N=100$, and (2) taking a conservative sum of our systematic and
statistical errors our findings are consistent with $\bv$ of unity at
the $\sim 2\sigma$ level.

\subsection{Redshift evolution}

\begin{figure}
\includegraphics[width=\columnwidth]{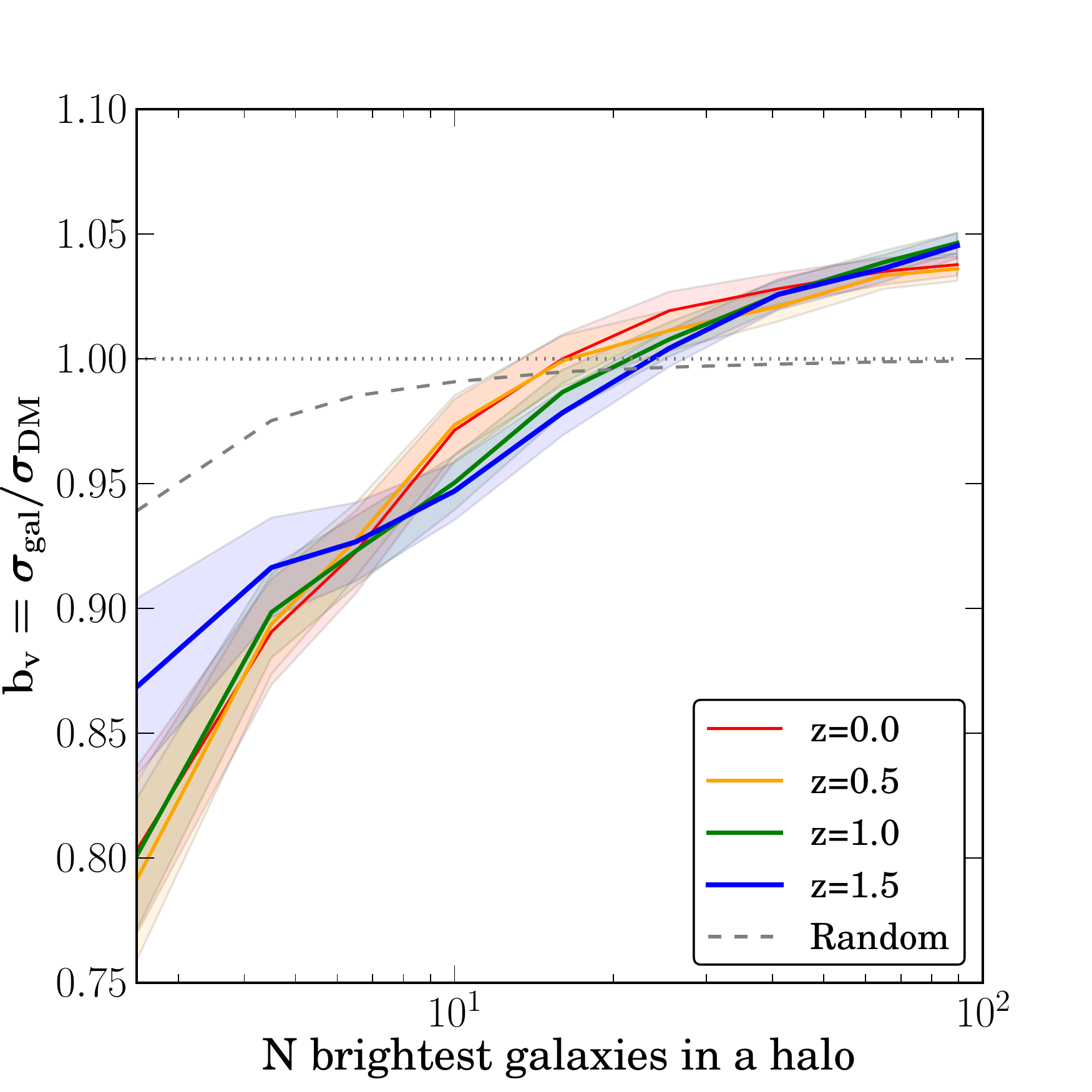}
\caption{Velocity bias of galaxies selected by $\vpk$ in the DM
simulation, for $z$ = 0, 0.5, 1, and 1.5.  For high $N$, the velocity
bias is consistent with no evolution.  For low $N$, low-redshift
bright galaxies are slightly slower than high-redshift ones,
indicating that the dynamical friction has been lasting for a longer
time.  }
\label{fig:redshift}
\end{figure}

Fig.~\ref{fig:redshift} shows the velocity bias at four different
redshifts from the {\sc rhapsody} simulation, using $\vpk$ as a
luminosity proxy, as a function of the number of brightest galaxies
used.  For high $N$, the results are independent of redshift.  For low
$N$, $\bv$ has a slight trend with redshift; the brightest galaxies at
low redshift tend to be slower than those at high redshift.  This is
plausibly explained by the effect of dynamical friction.  Since haloes
at different redshifts also have different masses, our finding here
implies that this relation is independent of mass.  We have explicitly
verified this statement.

\cite{Munari13} have shown that the velocity bias decreases at high
redshift, which is in apparent disagreement with our findings.  This
difference could result from the different galaxy population we use;
while we compare the same number of galaxies per halo across all
redshifts, \cite{Munari13} set constant thresholds for the dark mass
of subhaloes ($10^9\rm M_\odot$) and the stellar mass of galaxies
($3\times10^9\rm M_\odot$) across all redshifts.  This constant
threshold will lead to fewer galaxies per cluster at high redshift.
Since fewer galaxies can results in lower $\bv$, this constant
threshold can result in a smaller $\bv$ at high redshift.

\section{Virial scaling from cluster galaxies}\label{sec:rank}

\begin{figure}
\includegraphics[width=\columnwidth]{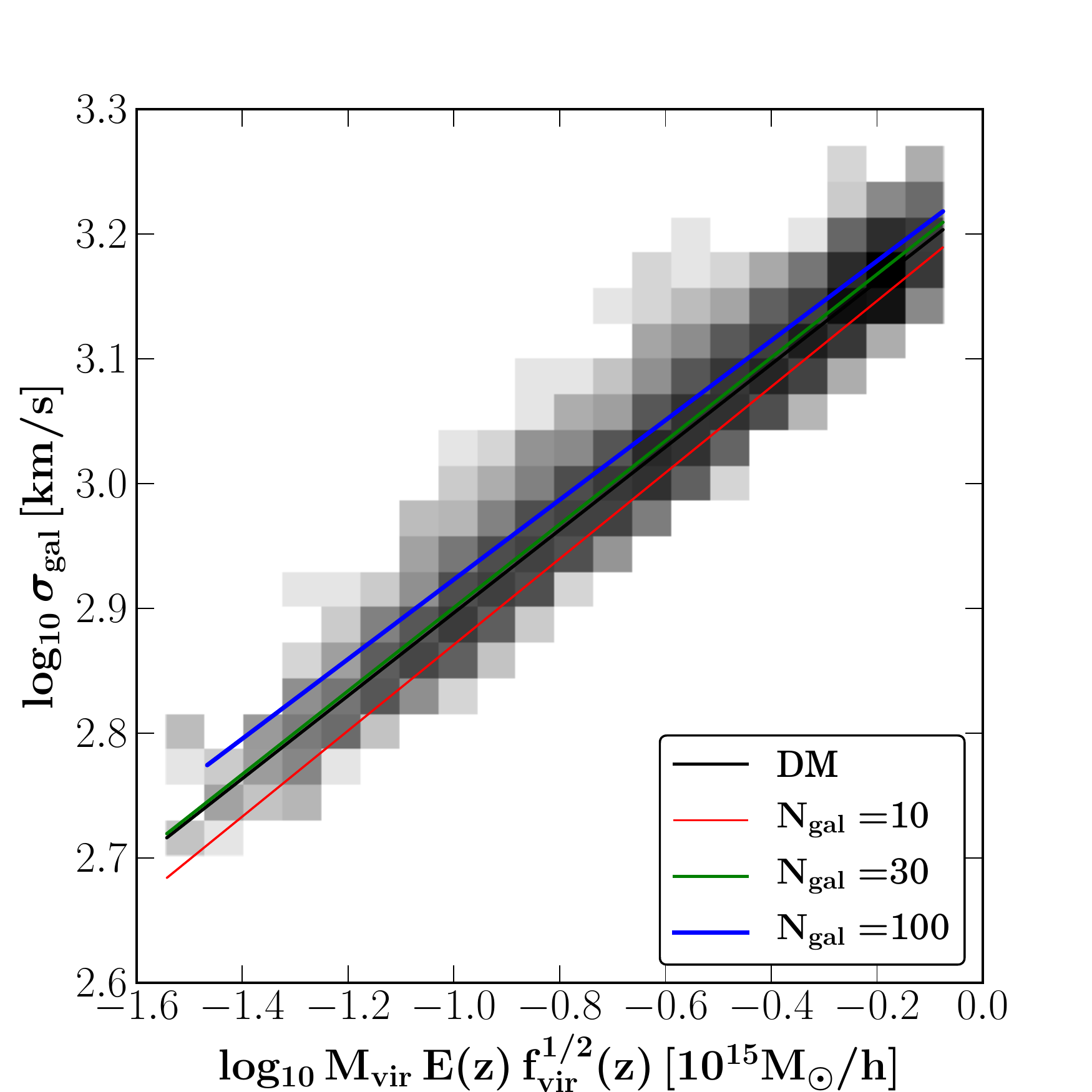}
\caption{The virial scaling relation calibrated using galaxies and DM
particles in {\sc rhapsody}.  The grey 2D histogram corresponds to
measurements from the DM particles, and the best-fitting relation
agrees with \protect\cite{Evrard08}.  The three colour curves
represent the fit of the virial scaling relation using the 10/30/100
brightest galaxies in each halo, and the values of fitting parameters
are given in Table~\ref{tab:M_sigma_fit}.  When we use the 10
brightest galaxies (red curve), the normalization is lower by $\approx
3\%$.  On the other hand, with 100 galaxies, the normalization is
higher by $\approx 3\%$.  }
\label{fig:M_sigma_fit}
\end{figure}
\begin{table}
\centering
\setlength{\tabcolsep}{0.5em}
\begin{tabular}{
>{\centering}p{1cm}<{\centering}  
>{\centering}p{1.5cm}<{\centering} 
>{\centering}p{1.5cm}<{\centering} 
>{\centering}p{1.5cm}<{\centering} 
c
}
\hline
\rule[-2mm]{0mm}{6mm}
$N_{\rm gal}$ & $\sigma_{\rm gal,15}$ & $\alpha$ & $\sqrt{\langle\delta_{\ln\sigma}^2\rangle}$ & shot noise\\ \hline 
\rule[-2mm]{0mm}{6mm} 10 &  1641$\pm$32 &   0.34$\pm$0.01 & 0.142 & 0.134 \\ \hline
\rule[-2mm]{0mm}{6mm} 30 &  1716$\pm$33 &   0.33$\pm$0.01 & 0.092 & 0.073 \\ \hline
\rule[-2mm]{0mm}{6mm} 100 &  1746$\pm$34 &   0.32$\pm$0.01 & 0.072 & 0.039 \\ \hline
\rule[-2mm]{0mm}{6mm} DM &  1692$\pm$33 &   0.33$\pm$0.01 & 0.053 & NA\\ \hline
\end{tabular}
\caption{Fitting parameters in Fig.~\ref{fig:M_sigma_fit}.
The scatter of each fit agrees with the intrinsic scatter combined with the shot noise.
}
\label{tab:M_sigma_fit}
\end{table}

Fig.~\ref{fig:M_sigma_fit} presents the virial scaling between cluster
mass and $\sigmagal$ measured with 10/30/100 brightest galaxies
(red/green/blue), compared with $\sigmaDM$ (black).  We use 85 time
steps between $z$ = 0 and 2 from the {\sc rhapsody} simulation and use
$\vpk$ as the luminosity proxy, assuming that different evolutionary
stages of the same haloes can provide a fair sample of the dynamical
states.  We justify this assumption by examining whether the virial
scaling ($\sigmaDM$--$\Mvir$ relation) of this sample agrees with
previous results from the volume-limited sample of
\citet[][]{Evrard08}.  The grey-scale is a 2D histogram of $\sigmaDM$
and $\Mvir$ of these haloes, whose virial masses are scaled with the
Hubble parameter $E(z) = h(z)/h_0$ and the square root of the virial
overdensity $f_{\rm vir} = \Delta_{\rm vir}(z)/\Delta_{\rm vir}(z=0)$,
for the purpose of eliminating the effect of the evolution of the
background density and the virial overdensity.  These haloes follow
the scaling relation
\beq
\ln\sigmaDM=\ln \left[1692.17\,\kms\right] + 0.33\, \ln \left[ \Mvir E(z) f^{1/2}_{\rm vir}(z) \right]
\eeq
with a scatter 
\beq
\langle\delta_{\ln\sigma}^2\rangle^{1/2} = 
\left\langle(\ln\sigma - \ln\sigma_{\rm fit})^2\right\rangle^{1/2} = 0.05
\eeq
These values agree with those quoted in \cite{Evrard08} (see their
table 6; normalization $982\sqrt{3}=1700$, slope 0.355, and scatter
0.0527), indicating that using multiple snapshots for each halo can
provide a fair sample of the dynamical state of
haloes.\footnote{Recently, \cite{Diemer13} have shown that this virial
scaling relation is robust regardless of the accretion rate,
indicating that haloes tend to be in local Jeans equilibrium.}

We now compare the virial scaling measured from different galaxy
samples.  The colour lines show the best-fitting linear relation, and
Table~\ref{tab:M_sigma_fit} lists the best-fitting parameters of the
following parametrization:
\beq
\ln\sigmagal = \ln\sigma_{\rm gal,15} + \alpha \ln \left[ \Mvir E(z) f^{1/2}_{\rm vir}(z) \right] 
\eeq
Using the 10 brightest galaxies to calibrate the scaling relation
biases the normalization low by $\approx3\%$.  When we use the 30
brightest galaxies, the normalization agrees almost perfectly with the
results using DM particles.  When we increase the number to 100, the
normalization is higher by $\approx3\%$.  This trend of normalization
with galaxy number agrees with the trend of velocity bias presented in
Fig.~\ref{fig:rank}.

We note that these different selections of galaxies tend to give the
same slope as using DM particles, $\approx$0.33.  This result is
contrary to the findings of \cite{Munari13}, who have found a slope of
$\approx 0.36$ for the virial scaling from subhaloes and galaxies.
Again, this discrepancy could result from the different subhalo/galaxy
population we use.  While we use a constant number of galaxies per
cluster, \cite{Munari13} set constant mass thresholds for subhaloes
and galaxies.  For less massive haloes, galaxies above these
thresholds are rare and relatively massive compared with the host
halo; thus, these galaxies are prone to the effect of dynamical
friction and become slower.  Therefore, a constant galaxy mass
threshold can cause $\bv<1$ for less massive haloes, thus leading to a
steeper slope.

Finally, we compare the scatter,
$\langle\delta_{\ln\sigma}^2\rangle^{1/2}$, of each fit.  We have
shown above that the virial scaling based on DMr particles presents a
5\% intrinsic scatter.  If we choose a small number of galaxies
$N_{\rm gal}$ to calculate the velocity dispersion, there is an
additional statistical error (``shot noise''), presented in the last
column of Table~\ref{fig:M_sigma_fit}.  We note that for the $N_{\rm
gal}$ we considered here (10/30/100), the total scatter
(0.14/0.09/0.07) is consistent with the combination of intrinsic
scatter (0.05) and shot noise (0.13/0.073/0.040).  Our values also
bracket the values quoted in \cite{Munari13} (their fig.~6).

We note that a 3\% bias in velocity dispersion corresponds to a 9\%
bias in dynamical mass, which can be a dangerous source of systematic
error if one uses spectroscopic follow-up for mass calibration for
photometric cluster surveys \citep[e.g.,][]{Wu09}.  In addition, an
accurate calibration of the scatter of the mass--observable
distribution is essential for accounting for the Eddington/Malmquist
bias of the cluster abundance and for interpreting the observed
massive, distant clusters \citep[e.g.,][]{Mortonson11}.

\section{Velocity structure of cluster galaxies}
\label{sec:bv_dm}

\subsection{Dependence of velocity bias on radius and luminosity: a stacking analysis}
\begin{figure}
\includegraphics[width=\columnwidth]{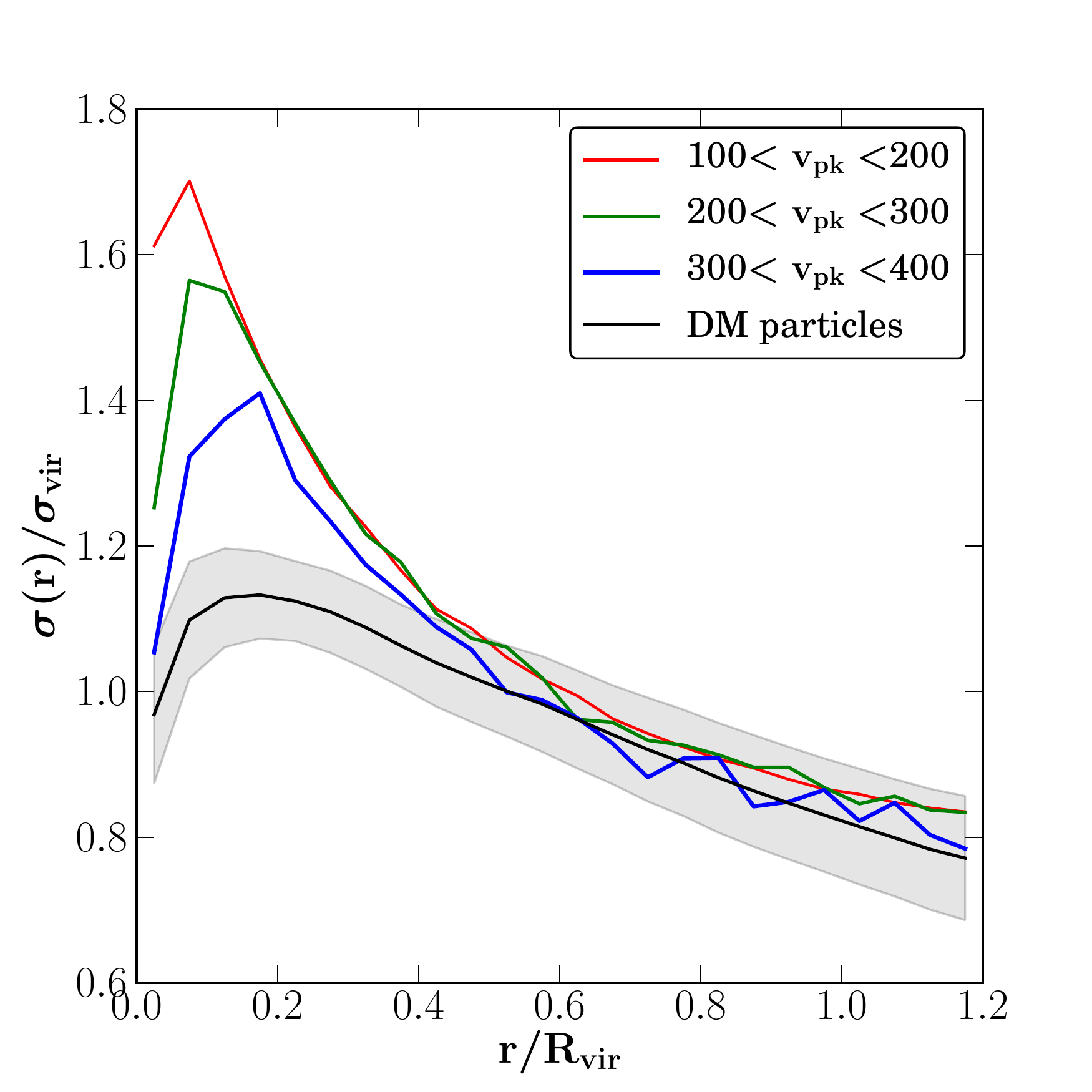}
\caption{Velocity dispersion profile for DM (black curve) and cluster
galaxies (colour curves) in {\sc rhapsody}.  Galaxies are binned by
the luminosity proxy $\vpk$.  Galaxies in all three bins have higher
velocity dispersion than DM particles within $0.5\Rvir$.  The red
curve, which corresponds to the faintest galaxies, has the highest
velocity bias.}
\label{fig:profile}
\end{figure}

\begin{figure}
\includegraphics[width=\columnwidth]{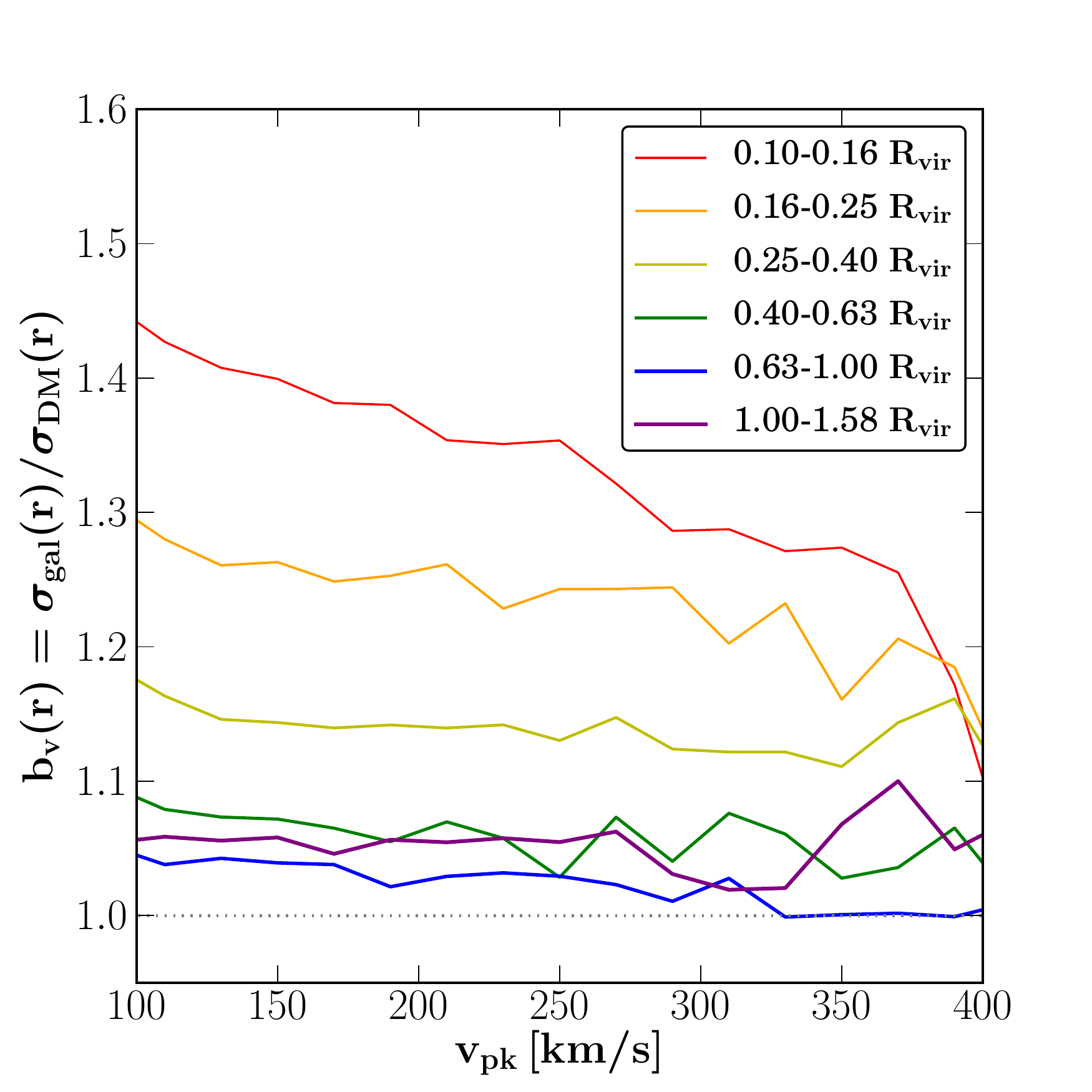}
\caption{Velocity bias for galaxies in {\sc rhapsody} as a function of
luminosity proxy $\vpk$, binned by radius (different curves).
Galaxies in the inner region of clusters (red and orange) tend to have
the highest velocity bias, which can be caused by strong tidal
stripping and disruption in this region.  Brighter galaxies (larger
$\vpk$ values) tend to be slower due to dynamical friction.}
\label{fig:radius}
\end{figure}

Two major physical processes that can produce velocity bias in
collisionless systems are dynamical friction and tidal disruption.
Dynamical friction tends to slow down galaxies and make them approach
$\bv\sim 1$ \citep[e.g.,][]{Boylan-Kolchin08}.  Tidal disruption tends
to occur in high density regions and remove slow galaxies
\citep[e.g.,][]{WetzelWhite10}, thus increasing velocity bias.  We
would like to disentangle these processes by examining the velocity
bias as a function of galaxy luminosity and the distance to the
cluster centre.  To this end, we use the rich statistics of cluster
galaxies in {\sc rhapsody} to study their velocity structures,
focusing on how the galaxy velocity depends on luminosity (modelled by
$\vpk$) and its distance to cluster center.  We use 32 outputs between
$0\leq z\leq 0.5$ and assume that the relation between luminosity and
$\vpk$ has negligible evolution in this redshift range.

Our stacking process is as follows. For a galaxy with velocity ${\bm
v_{\rm gal}}$ in a halo of velocity dispersion $\sigmavir$, we define
\beq
{\bm v'} = \frac{{\bm v_{\rm gal}} - {\bm v_{\rm cen}}}{\sigmavir} \ ,
\eeq
where $\bm v_{\rm cen}$ is the velocity of the BCG, and $\sigmavir$ is the 
DM velocity dispersion of the host halo.
The stacked velocity bias is given by
\beq
b_{\rm v,\,stack}^2 = \frac{1}{N-1}\sum_{i=1}^{N} || {\bm v'}_i - \bar{\bm v'} ||^2 \ ,
\eeq
where $N$ is the number of galaxies, and $\bar{\bm v'}$ is the mean of
${\bm v'}$ in this stack of galaxies.  This is analogous to
Equation~\ref{eq:sigma} but takes into account the different masses of
host haloes by dividing galaxy velocities by $\sigmavir$.  Under this
definition, the stacked velocity bias is the dispersion of velocity
ratio, rather than the ratio of velocity dispersion.

Fig.~\ref{fig:profile} presents the stacked velocity bias profile of
galaxies, binned by the luminosity proxy $\vpk$ (colour curves).  The
velocity dispersion profile of DM is presented as the black curve.
Beyond 0.5$\Rvir$, galaxy velocity is roughly unbiased.  The brightest
galaxies (blue) has slightly lower velocity bias, indicating that the
effect of dynamical friction is stronger for brighter/more massive
galaxies.  On the other hand, at smaller radii, the faintest sample
(red) shows the largest velocity bias, while the brightest sample
shows the least velocity bias.  This can be explained by the tidal
disruption of faint galaxies, which is stronger at smaller radii,
where the density is high.  We note that \cite{Ludlow09} have shown
that in Milky Way-size haloes, the radial velocity dispersion profile
of subhaloes depends on subhalo mass; low-mass subhaloes have higher
velocity bias than high-mass subhaloes.  This trend is consistent with
our findings.

Fig.~\ref{fig:radius} shows another aspect of the radial and
luminosity dependence of velocity bias.  Galaxies are put into
logarithmically spaced radial bins (5 bins per decade), as indicated
by the legend.  To focus on the {\em local} velocity bias, in the
stacking process, galaxies in a radial bin are compared with the DM in
the same radial bin; that is,
\beq
{\bm v'}(r) = \frac{{\bm v_{\rm gal}} - {\bm v_{\rm cen}}}{\sigmaDM(r)} 
\eeq
The $x$-axis indicates the luminosity proxy $\vpk$, while the $y$-axis
indicates the stacked velocity bias in for a given radial bin.

For faint galaxies (low $\vpk$), the inner radial bins (red, orange,
and yellow) show significant velocity bias with a clear trend with
radius.  At small radii, faint galaxies experience stronger tidal
stripping and are more easily disrupted, and the surviving subhaloes
tend to be the fast ones and have high velocity bias.  In addition,
bright galaxies (high $\vpk$) tend to be slower than faint galaxies
because they have been slowed down more by dynamical friction.  At
large radii, the dependence on luminosity is weak, because galaxies
are still infalling and have not experienced dynamical friction for
long.

\subsection{Velocity distribution function of cluster galaxies}
\begin{figure}
\includegraphics[width=\columnwidth]{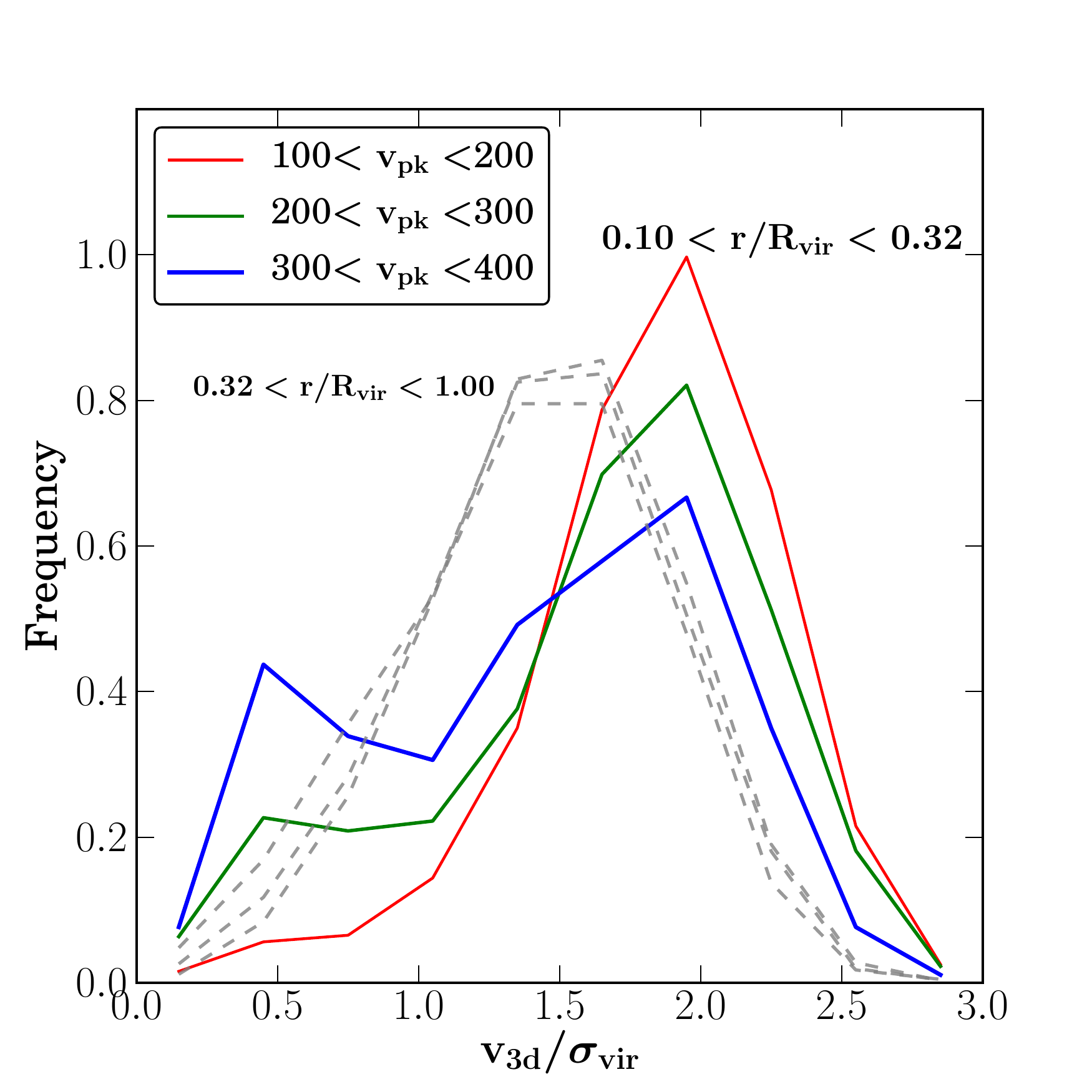}
\caption{The velocity distribution function for different galaxy
populations.  Here we highlight $0.1<r/\Rvir<0.32$, and different
colour curves present different luminosity bins (modelled by $\vpk$).
Bright galaxies have a highly negatively-skewed VDF, with a heavy tail
of slow galaxies.  This trend indicates that the velocity dispersion
inferred from bright galaxies near the cluster centre tend to be
lower.  We note that this trend is largely diminished at larger radii
(grey dashed curves correspond to $0.32<r/\Rvir<1$).}
\label{fig:VDF}
\end{figure}

In addition to the velocity dispersion, we would like to quantify the
full velocity distribution function (VDF) of cluster galaxies to
improve our understanding of the velocity structure of different
galaxy populations.  Fig.~\ref{fig:VDF} presents the VDF of cluster
galaxies from stacking the {\sc rhapsody} sample ($0\leq z\leq 0.5$),
using ${\bm v'}$ as the scaled velocity of each galaxy.  Galaxies are
binned by the luminosity proxy $\vpk$.

We have found that most radial ranges give very similar VDFs (an
example is shown by the grey dashed curves), except for the radial
range of $0.1 < r/\Rvir < 0.32$ (colour curves).  At this radial
range, bright galaxies show a highly negatively-skewed VDF, with the
presence of a slow population, reflecting the strong effect of
dynamical friction in this regime.  We note that this VDF tail is not
caused by the BCG, which is excluded in this analysis.  This trend
does not exist at larger radii; for comparison, the grey cash curves
show the radial range of $0.32 < r/\Rvir < 1$, where different
luminosity bins show similar VDFs.

Although this slow and luminous galaxy population is identified using
galaxy tracers in $N$-body simulations, we note that this population
also exists in hydro simulations, in which gas can further slow down
these galaxies.  This prediction is also robust with increased
resolution, because this galaxy population corresponds to
well-resolved subhaloes in the current simulation.

Finally, we caution that this slow population could potentially bias
the dynamical mass measurements.  When following up a cluster using
spectroscopy, one tends to start from the brightest galaxies closest
to the centre to reduce the impact of interlopers.  Our results
indicate that this strategy could bias the dynamical mass low.  In
addition to cluster mass calibration, we note that the VDF is also an
important input in the Jeans analysis \citep[e.g.,][]{Mamon13} and in
the modelling of redshift-space distortions of small-scale galaxy
clustering \citep[e.g.,][]{Tinker07}.  Therefore, an accurate
characterization of the cluster galaxy VDF is essential for both
cluster counting and galaxy clustering experiments.

\section{Comparing dark matter and hydrodynamical simulations}
\label{sec:bv_hydro}
\begin{figure}
\includegraphics[width=\columnwidth]{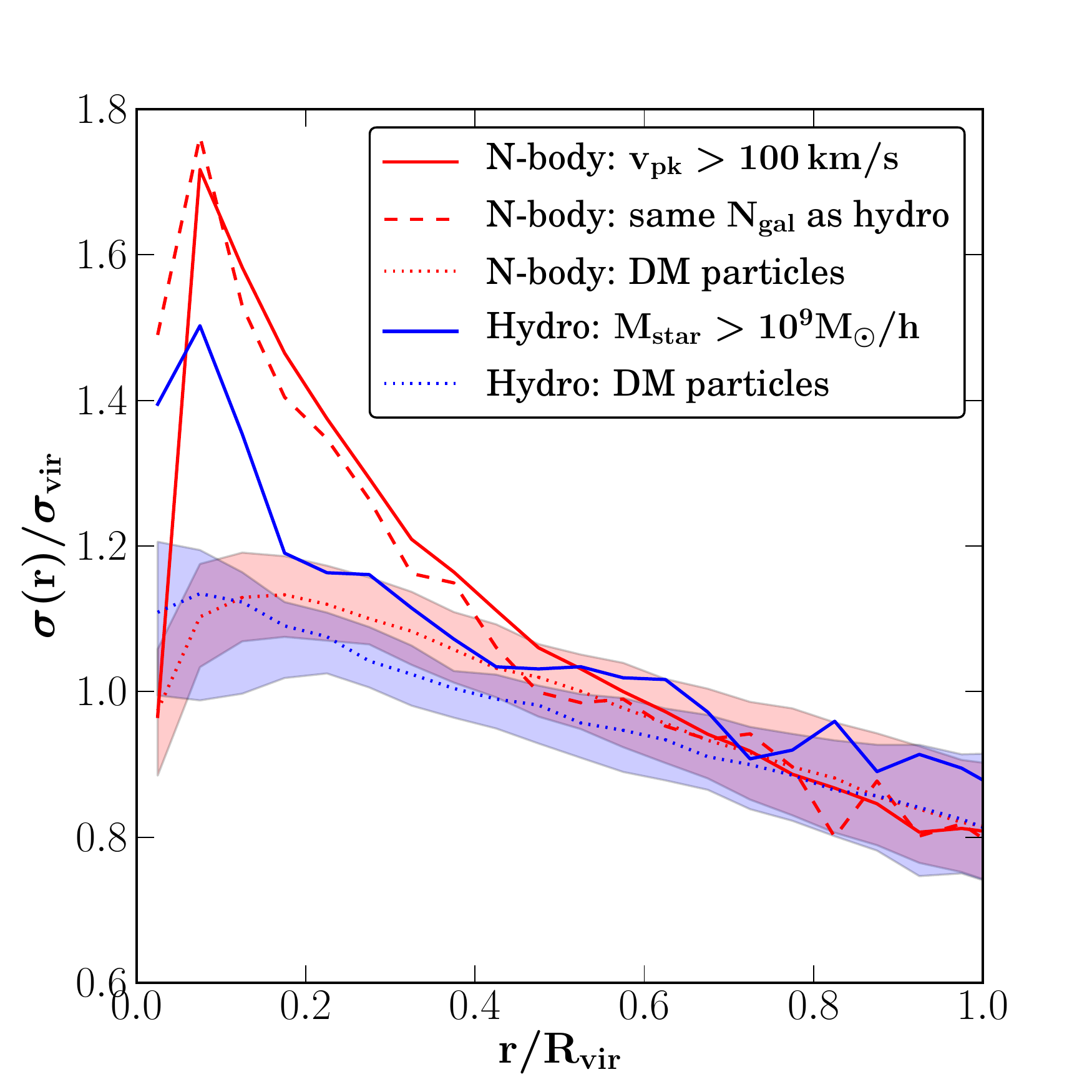}
\caption{Velocity dispersion profile of haloes from the $N$-body
simulation (red) and the hydro simulation (blue) at $z=0$.  Dark
matter particles in the hydro simulation tend to be faster at small
radii (dotted curves).  We note that this difference cannot be
accounted for by the mass differences in these two samples. On the
other hand, galaxies in the hydro simulation tend to be slower (solid
curves).  Overall, galaxies in the hydro simulation have smaller
velocity bias in the inner regions than those in the $N$-body
simulation.}
\label{fig:profile_hydro}
\end{figure}

In this section, we discuss how the velocity structure of cluster
galaxies could be altered by baryonic physics.  We compare the DM
simulation {\sc rhapsody} ($z=0$ sample) with the hydro simulation
{\sc magneticum pathfinder} introduced in Section~\ref{sec:sims} in
Table~\ref{tab:sims}.

Fig.~\ref{fig:profile_hydro} compares the velocity dispersion profile
of galaxies and DM particles in the $N$-body (red) and hydro (blue)
simulations.  Let us first compare the DM particles (dotted curves).
In the $N$-body simulation, the velocity dispersion decreases towards
the centre, which is expected from the Navarro--Frenk--White (NFW)
profile and is consistent with results from previous $N$-body
simulations \cite[e.g.,][]{ColeLacey96,Colin00,Navarro10}.  On the
other hand, in the hydro simulation, the velocity dispersion stays
nearly constant towards the center.  This flat velocity dispersion
profile has also been seen in the BCGs in the {\sc magneticum
pathfinder} simulation, as well as for elliptical galaxies in several
hydro simulations, in \cite{Remus13}.  Possible explanations of this
difference include (1) change in the density profile, and (2) change
in the kinematics with the implementation of baryonic physics.  When
we investigate the density profile, we find that the hydro simulation
does not produce an artificial high-density core (which has been seen
in CSF simulations), because the simulation includes the AGN feedback
and avoids the overcooling problem (see
e.g.,~\citealt{Teyssier11,Martizzi12} on the effect of AGN feedback on
the density profile).  \cite{Newman13} have also shown that hydro
simulations with AGN feedback can reproduce the observed cluster
density profiles better than those without.  Therefore, the
near-isothermal core is not due to a significant deviation of density
in the core.  On the other hand, the AGN feedback may increase the
kinetic energy of the DM particles near the cluster centre and make it
close to isothermal.  Early work without AGN feedback does not show
such a nearly isothermal velocity dispersion profile (see
e.g.~\citealt{Rasia04} for non-radiative hydrodynamical
simulations). This clearly presents a question that deserves further
investigation in the future.

We now turn to the comparison of the velocity dispersion of cluster
galaxies.  The blue solid curve corresponds to galaxies in the hydro
simulation with $\Mstar>10^9\hiMsun$, and the red solid curve
corresponds to galaxies in the $N$-body simulation with
$\vpk>100\,\kms$; this choice is based on abundance matching between
$\vpk$ and $\Mstar$ using cosmological simulations (Behroozi, private
communication).  To avoid the effect of the different galaxy number
density profiles between the $N$-body and hydro simulations, we add
another selection in the $N$-body simulation: the red dashed curve
corresponds to selecting the same number of galaxies as in the hydro
simulation at each radius.  At smaller radii ($r < 0.5\Rvir$),
galaxies in the DM simulation tend to be faster than those in the
hydro simulation, which can be attributed to stronger tidal disruption
in $N$-body simulations.  In addition, faint galaxies are dominant in
number in our sample and are more susceptible to tidal disruption.
Interestingly, in the innermost region ($r<0.1\Rvir$), the hydro
simulation has faster DM particles but slower subhaloes compared with
the $N$-body simulation in this region.  We note that the shaded
region indicates the halo-to-halo scatter (68 per cent
interval). Although the scatter is large, the mean is robustly
determined given our sample size.  Therefore, the difference in the
mean is significant.

We can see that the velocity bias is smaller in hydro simulations,
i.e., the velocity of galaxies follow that of DM more closely.  This
can be attributed to two effects.  First, in hydro simulations, the
gas can exert a dragging force on to the galaxies that slows them down
\citep[e.g.,][]{Puchwein05}.  Secondly, in hydro simulations, the
cores of subhaloes are denser because of cooling and subsequent star
formation in the center; therefore, these subhaloes tend to survive
longer than those in pure DM simulations.  These longer-lived
subhaloes are also the slow population, which contributes to an
overall lower velocity dispersion.

The second effect discussed above is related to the issue that
subhaloes in pure DM simulations are often incomplete due to
``over-merging'' when compared with subhaloes in hydrodynamical
simulations with star formation
\citep[e.g.,][]{Klypin99,Weinberg08,Dolag09}.  In semi-analytic galaxy
formation models, galaxies can be traced after subhalos disappear,
either due to this over-merging effect or just due to the finite
resolution; these are generally known as ``orphan galaxies''
\citep[e.g.,][]{Gao04}. Whether this over-merging problem has been
completely resolved by either hydro simulations or SAMs is still an
open question.  Future implementations of hydro simulations and galaxy
tracer models will need to be critically compared with and constrained
by observations, including the observed satellite content and radial
profiles of galaxies in groups and clusters
\citep[e.g.,][]{Budzynski12,Tinker12,Newman13,Reddick13}.

\section{Summary}\label{sec:summary}

Using $N$-body and hydrodynamical simulations of galaxy clusters, we
have studied the velocity bias of different galaxy tracers and the
virial scaling calibrated from galaxies.  In particular, we have used
the rich statistics of subhaloes in the {\sc rhapsody} $N$-body
simulation in parallel with galaxies identified in the {\sc magneticum
pathfinder} hydrodynamical simulation to study the velocity structure
of cluster galaxies.  Our findings can be summarized as follows.

\begin{itemize}

\item Galaxy populations inferred from either $N$-body simulations
with $\vpk$ as a luminosity proxy or hydrodynamical simulations show
that the brightest cluster galaxies tend to underestimate, and faint
galaxies slightly overestimate, the DM velocity dispersion.  Our
results indicate that selecting $\sim30$ brightest cluster galaxies
provide an approximately unbiased velocity dispersion.

\item Ranking subhaloes by their instantaneous circular velocity,
$\vtoday$, in $N$-body simulations leads to a larger velocity bias
incompatible with the other galaxy tracers.  We demonstrate
consistency of our results with results in the literature that employ
different tracer methods.

\item Velocity bias tends to be higher for fainter galaxies at smaller
cluster radii, indicating the stronger effect of tidal disruption,
which removes slow galaxies efficiently in high density regions.
Brighter galaxies are in general cooler, and we have identified that
the brightest galaxies near the centre of a cluster (not the BCG)
present a particularly slow population, which is plausibly consistent
with dynamical friction.

\item Comparing the hydrodynamical and $N$-body simulations, we have
found that galaxy tracers in the former are less kinematically biased
with respect to the DM.  At small cluster radii, the DM in the hydro
simulation has higher velocity dispersion than that in the $N$-body
simulation, while the galaxies in the hydro simulations have {\em
lower} velocity dispersion.

\end{itemize}

Current and near-future observed cluster samples with dense
spectroscopy will be able to test for the differential effect from
bright to faint magnitudes expected from simulations.  Stacked sample
analysis, such as has been done recently by \citet{Skielboe12} to
detect spatial velocity anisotropy, offers an attractive method to
boost signal-to-noise ratio.

We remark that understanding the {\em intrinsic} dynamical properties
of simulated galaxy tracers is important for using simulations to
study observational complications such as projection effects and
member galaxy selection \citep[e.g.,][]{Gifford13,Gifford13b,Saro12}.
In addition, the velocities of galaxies are essential for modelling
the small-scale redshift-space distortions in galaxy redshift surveys.
An accurate calibration of velocity bias will improve the usage of
small-scale clustering information to constrain cosmological
parameters \citep[e.g,][]{WuHuterer13,ZuWeinberg13}.

\section*{Acknowledgements}
We thank Dragan Huterer, Elena Rasia, Chris Miller, and Dan Gifford
for helpful discussions, and Peter Behroozi for providing the
relationship between $\Mstar$ and $\vpk$ for the model used here.  We
also thank Eduardo Rozo, Erwin Lau, Raul Angulo, and the anonymous
referee for helpful suggestions.  HW acknowledges the support by the
US Department of Energy under contract number DE-FG02-95ER40899.  OH
acknowledges support from the Swiss National Science Foundation (SNSF)
through the Ambizione fellowship.  KD acknowledges the support by the
DFG Cluster of Excellence ``Origin and Structure of the Universe.''
This work received support from programme number HST-AR-12650.01-A,
provided by NASA through a grant from the Space Telescope Science
Institute, which is operated by the Association of Universities for
Research in Astronomy, Inc., under NASA contract NAS5-26555.  The {\sc
rhapsody} simulations were run using computational resources at SLAC
and supported by the U.S.\ Department of Energy under contract
DE-AC02-76SF00515 to SLAC.

\bibliographystyle{mn2e}
\bibliography{master_refs}

\end{document}